\documentclass[prd,showpacs,preprintnumbers,floatfix,superscriptaddress,11pt]{revtex4}
\usepackage[mathscr]{eucal}
\usepackage{color}
\usepackage[dvips]{graphicx}
\usepackage{epsf}
\usepackage{bm}
\usepackage{epsfig}
\usepackage{feynmf}
\usepackage{amsmath}

\newcommand{\be}{\begin{equation}}
\newcommand{\ee}{\end{equation}}
\newcommand{\ba}{\begin{eqnarray}}
\newcommand{\ea}{\end{eqnarray}}

\newcommand{\pa}{\partial}

\newcommand{\Om}{\mathbf{\Omega_p}}
\newcommand{\Evec}{\mathbf{E}}
\newcommand{\Bvec}{\mathbf{B}}

\begin{document}

\title{Kinetic theory of chiral relativistic plasmas \\ and energy density of their gauge collective 
excitations}

\author{Cristina Manuel}
\email{cmanuel@ieec.uab.es}
\affiliation{Instituto de Ciencias del Espacio (IEEC/CSIC) Campus Universitat Aut\`onoma de Barcelona, Facultat de Ci\`encies, Torre C5, E-08193 Bellaterra (Barcelona), Catalonia, Spain}
\author{Juan M. Torres-Rincon}
\email{jtorres@ieec.uab.es}
\affiliation{Instituto de Ciencias del Espacio (IEEC/CSIC) Campus Universitat Aut\`onoma de Barcelona, Facultat de Ci\`encies, Torre C5, E-08193 Bellaterra (Barcelona), Catalonia, Spain}


\begin{abstract}

We use the recently developed  kinetic theory with Berry curvature to 
describe the fermions and antifermions of a chiral relativistic plasma.
We check that this transport approach allows to reproduce the chiral anomaly equation of 
relativistic quantum field theory  at finite temperature.
We also check that it allows to describe the anomalous gauge 
polarization tensor that appears in the Hard Thermal (and/or Dense)
effective field theory. We  also construct an energy density associated to 
the gauge collective modes of the chiral relativistic  plasma, valid in 
the case of small couplings or weak fields, which can be the basis for the study 
of their dynamical evolution.
\end{abstract}

\date{\today}
\maketitle

\section{Introduction}

It has been suggested that in the early stages of heavy-ion collisions 
there should be a chirality imbalance between
left- and right-handed quarks, whose origin is the event-by-event fluctuations of the topological charge due to the excitations
of the gluonic configurations~\cite{Kharzeev:2004ey,Kharzeev:2007jp,Kharzeev:2009fn}.
In the presence of a magnetic field created by the colliding ions this 
chiral imbalance would lead  to a current of electric charge, the so-called chiral
magnetic effect (CME)~\cite{Fukushima:2008xe}. Such an 
effect, noted earlier in different
physical scenarios~\cite{Vilenkin:1980fu,Nielsen:1983rb},
is related to the quantum chiral anomaly \cite{Adler,BellJackiw}. The CME would lead to charge asymmetry
fluctuations in heavy ion collisions~\cite{Kharzeev:2010gr}. While there is a
reported charge asymmetry fluctuation found at the Relativistic Heavy-Ion Collider (RHIC)~\cite{Abelev:2009ac,Abelev:2009ad}, the interpretation of the phenomena 
is still not clear.

It has been recently shown that the effects of the  chiral anomalies 
that appear in relativistic Fermi liquids can be obtained in the 
framework of (semi) classical kinetic theory~
\cite{Son:2012wh,Stephanov:2012ki,Son:2012zy,Chen:2012ca}.
Much of the relevant work leading to that conclusion was done for 
condensed matter systems~\cite{Wong:2011nt,Shindou}, after analyzing the 
effect of the Berry phase~\cite{Berry:1984jv} --the phase that the 
fermion wavefunction picks up after performing a closed path in the 
Hamiltonian parameter space-- on the particle's dynamics. 
 In the 
presence of the Berry curvature Liouville's theorem and Hamiltonian 
dynamics are only possible after a modification of the phase space 
element and the  classical equations of motion~\cite{Duval:2005vn}. For 
a chiral fermion the modified action is one with a fictitious magnetic monopole field in 
momentum space. Such a monopole field, placed in the position where  the 
level crossing responsible of the chiral anomaly takes place, is 
the source of the non-conservation of the chiral current.

Classical or semi-classical transport equations for relativistic systems 
can be obtained from the  the underlying quantum field theory after 
defining a Wigner function  and performing a derivative expansion. 
Following this path, it has been shown that the gauge polarization 
tensor obtained from quantum field theory and with the transport equation of Ref.~\cite{Son:2012zy} agree, both in the parity even side~\cite{Manuel:1995td}, as 
well as in the parity odd part describing the CME effect, at zero 
temperature $T=0$~\cite{Son:2012zy}. This was done using the so-called 
High Density Effective Theory~\cite{Hong:1998tn}, an effective field 
theory valid for the fermionic modes near to the Fermi surface, and 
where other modes, including the antiparticles, are integrated out.
We should also stress here that most part of the literature on this 
topic claims that the kinetic equation with Berry curvature leads to  
the proper description of the chiral anomalies when there is a 
well-defined Fermi surface (that is, at finite density and $T=0$)
~\cite{Son:2012wh,Stephanov:2012ki,Son:2012zy,Chen:2012ca,Basar:2013qia}.
Only in a footnote of Ref.~\cite{Stephanov:2012ki} it was mentioned how to properly take into
account thermal effects.
  In this work   we also show that the agreement also occurs at any 
finite value of the temperature  $T$.
 We first check that in thermal 
equilibrium one can obtain the proper chiral anomaly equation, if we 
also consider a transport equation with Berry curvature for the 
antifermions, although we do not attempt to derive such a transport 
equation from the underlying quantum field theory. We also see that if we extend the formulation of the chiral transport equation of
Ref.~\cite{Son:2012zy} at finite $T$ including antifermions we can reproduce the induced current and associated effective action obtained  
after computing the anomalous Feynman diagrams in an Abelian theory when 
a chemical potential is assigned to chiral fermions
~\cite{Redlich:1984md,Laine:2005bt}. We should warn here that
although the different formulations 
 of the chiral transport theory available in the literature look equivalent, they are not. They differ in the form of
 the fermion dispersion law, and in  terms that appear in the transport equation, as
 we will point out later on this manuscript.

The use of transport theory has shown to be very fruitful for the study 
of different sort of relativistic plasmas~\cite{Blaizot:2001nr,Litim:2001db}. It allows for the study of 
purely out-of-equilibrium phenomena, and thus, of  the dynamical 
evolution of the system.
It describes properly the close-to-equilibrium domain of the plasma, 
reaching to the same sort of answers as obtained from quantum field 
theory for long (or soft) scales, with ease, while it allows for a simple
numerical implementation.  In this manuscript we also prove the 
versatility of transport theory by obtaining
an energy density describing the gauge collective modes of a chiral 
relativistic Abelian plasma. This energy density is given in local form, 
and can be the basis for numerical studies of the time evolution of the
CME in the cases where the gauge coupling constant is small or for weak 
fields.
  To achieve that goal, we simply generalize the same procedure used to 
obtain the energy density for the gauge collective modes of a weakly 
coupled quark-gluon plasma of Ref.~\cite{Blaizot:1994am}.

  This paper is organized as follows.  In Sec.~\ref{aHTLeffectiveaction}  we review how in
an Abelian gauge field theory with a chirality imbalance a Chern-Simons term
is generated in the static limit, after integrating out the high fermionic momentum modes. In the non-static limit, a much more complicated non-local structure is generated instead. We give the expressions of the effective actions in both cases, as well as
the value of the corresponding gauge field polarization  tensor. In Sec.~\ref{sec:kintheory}   we present a brief
summary of the recently proposed kinetic theory with Berry curvature corrections, see Ref.~\cite{Son:2012zy}.
We emphasize the fact that in order to reproduce properly the chiral anomaly equation of a relativistic quantum field theory at finite temperature both fermions and antifermions have to be
described on similar terms. In Sec.~\ref{sec:lrt} we use linear response theory to show how the proposed kinetic theory of Ref.~\cite{Son:2012zy} reproduces the results presented in Sec.~\ref{aHTLeffectiveaction}.
In Sec.~\ref{sec:hamiltonian} we construct an energy density valid to describe the collective modes of
the chiral plasma, at the order of accuracy we used to solve the transport equations in Sec.~\ref{sec:lrt} . In Sec.~\ref{sec:CSwave} we present a static solution to the gauge field
equations, together with the value of its energy density. Sec.~\ref{sec:conclusions} is devoted to summarize our main findings and conclusions.
Throughout the paper we use natural units, $k_B = c = \hbar =1$.

\section{Anomalous Hard Thermal Loop effective action}
\label{aHTLeffectiveaction}

It has been known for a while that a Chern-Simons term is induced for gauge fields in a theory where there is a finite
chemical potential assigned to the chiral fermions~\cite{Redlich:1984md}. While in the static limit the Chern-Simons term is local, a  non-local structure 
is however generated in the non-static situation.
After an analysis of the  Feynman diagrams, the corresponding effective
action has been derived in  Ref.~\cite{Laine:2005bt}. The proper way to obtain the effective action is in the framework of
the Hard Thermal (or Dense) Loop theory (HTL/HDL)~\cite{Braaten:1989mz,Frenkel:1989br,Manuel:1995td}. Then, one integrates out the the ``hard" fermions, that is fermions with energy
of the order $T\gg \mu$ (or $\mu\gg T$), and  then  obtains an effective 
field theory for the ``soft" gauge fields, that is, or order $gT$ for  $T \gg \mu$ (or $g\mu$ for $\mu \gg T$).  

While the results of Ref.~\cite{Laine:2005bt} were obtained for hypermagnetic fields, as their coupling with fermions is chiral, they can be generalized to
any other Abelian theory.
For definiteness, let us consider the case of a $U(1)$ gauge symmetry, coupled to massless {\it right-handed} fermionic fields with a chemical potential $\mu_R$ and {\it left-handed} fermionic fields with a chemical potential $\mu_L$.
The gauge field strength is $F_{\mu \nu}(x) = \pa_\mu A_\nu(x) - \pa_\nu A_\mu(x)$, expressed in terms of the vector gauge field $A_\mu(x)$.
After integrating out the hard fermionic  degrees of freedom, one gets the standard parity even (+) HTL effective action~\cite{Braaten:1991gm}~\footnote{ To recover the results of Ref.~\cite{Laine:2005bt}, which were derived for hyperelectromagnetic fields,  we should make the identifications $g'/2 = e$ and $Y=1$.
Note also that we include a factor of $1/2$ in the definition of the dual field strength, different from the definition  used in Ref.~\cite{Laine:2005bt}},

\be 
S^+_{\rm HTL} =  - \frac{m^2_D}{4} \int_{x,v} \ F_{\alpha \mu}(x) \ \frac{v^\alpha v^\beta}{(v\cdot \partial)^2} \ F_{\beta}^{\ \mu} (x)   \ , 
\ee
where 
\be
m^2_D = e^2 \left(\frac{T^2}{3} + \frac{\mu_R^2 +\mu_L^2}{2 \pi^2} \right)
\ee
is the Debye mass, and $v^\mu \equiv (1 ,{\bf v})$ where ${\bf v}$ is a unit vector. We have introduced the shorthand notations
\be
\int_x  \equiv \int d^4 x \ , \qquad
\int_v  \equiv \int \frac{d \Omega_v}{4 \pi}  \ ,
\ee
with $d\Omega_v$ the solid angle element.

The anomalous, parity odd (--), HTL action can be expressed as \cite{Laine:2005bt}
\be 
S^-_{\rm HTL} = \frac{c_E e^2}{4 \pi^2} \ \int_{x,v} \ \widetilde{F}_{\alpha \mu}(x) \ \frac{v^\alpha v^\gamma}{(v \cdot \pa)^3} \ F^{\ \mu}_{\gamma}(x) \ , 
\ee
where $\widetilde {F}_{\alpha \beta}(x) = \frac{1}{2} \epsilon_{\alpha \beta \mu \nu} F^{\mu \nu} (x)$ is the dual field strength and
$c_E = -\mu_5/2$, with $\mu_5 = \mu_R- \mu_L$. 

In the static limit ($\pa_0 \ll \pa_i$) it becomes the (local) Chern-Simons action~\cite{Redlich:1984md}

\be
S^-_{\rm{HTL}} = - \frac{c_E e^2}{8\pi^2} \int_{x} \ \epsilon_{ijk} A_i(x) F_{jk}(x) \ . 
\ee

Since the above effective actions are quadratic in the gauge fields, their content can be equivalently expressed through equations of motion.
In this case, one obtains a Maxwell equation with a source~\cite{Laine:2005bt}
\be
\label{anomalousMaxwell}
 \pa^\mu F_{\mu \nu} (x)  = - \int_v \left[   m^2_D  \frac{v_\nu v^\alpha }{(v \cdot \pa)} F_{\alpha 0}(x) + \frac{c_E e^2}{2 \pi^2} \epsilon_{0 \nu \alpha \beta} \frac{ v^\alpha }{(v \cdot \pa)} \pa^2 A^\beta (x)   \right] \ ,
\ee
where $\pa^2 \equiv \pa_\mu \pa^\mu$.
Here, and different than in Ref.~\cite{Laine:2005bt}, we assume that the system is electrically neutral, which can be achieved with a background static electric charge.
Therefore, a field-independent term in the current is absent, as opposed to what happens in Ref.~\cite{Laine:2005bt}.

In the static limit, the above equations indicate that there is an electromagnetic current proportional to the magnetic field
\be \label{eq:CME}
{\bf J}(x) = \frac{e^2 \mu_5}{4 \pi^2} \, {\bf B}(x) \ .
\ee
This is the so-called chiral magnetic effect, recently discussed in the literature in the framework of heavy-ion collisions~\cite{Kharzeev:2004ey, Kharzeev:2007jp, Fukushima:2008xe,Kharzeev:2009fn} and
in holographic models~\cite{Yee:2009vw,Rebhan:2009vc,Gynther:2010ed}.

It is also possible to derive the gauge field polarization tensor from the effective action. One obtains
\be
\Pi^{\mu \nu} (k) =   \Pi_+^{\mu \nu} (k) +   \Pi_-^{\mu \nu} (k) \ ,
\ee
where $k^\mu = (\omega, {\bf k})$, 
and is composed by the standard HTL, parity even  part
\be 
\label{pol-HTL}
 \Pi_+^{\mu \nu} (k)= - m^2_D \left(  \delta^{ \mu0 } \delta^{ \nu 0} -  \omega \int_v \  \frac{v^{\mu} v^{\nu}}{v \cdot k}  \right) \ ,  \ee
and the anomalous HTL, or parity odd part
\be
\label{pol-aHTL}
 \Pi_-^{\mu \nu} (k) =  \frac {c_E e^2}{ 2 \pi^2} i \epsilon^{\mu \nu \alpha \beta } k^2 k_\beta \int_v \frac{v_\alpha}{(v \cdot k)^2} \ .
\ee
Retarded boundary conditions are assumed in the expressions of the polarization tensors, with the prescription $\omega \rightarrow \omega + i 0^+$.
Note that the polarization tensor obeys the Ward identity $k_\mu \Pi^{\mu \nu} (k) =0$, and thus, it is compatible with gauge invariance.
This implies that the electromagnetic current that appears as a source of the Maxwell equation (\ref{anomalousMaxwell}) is conserved.

Explicit expressions for $\Pi^{\mu \nu}_+$, obtained after performing the angular  integrals in ${\bf v}$ can be found for example in Ref.~\cite{Litim:2001db}.
One can evaluate $\Pi^{\mu \nu}_-$ to find out that $\Pi^{\mu 0}_-=\Pi^{0 \mu}_- =0$, while

\be
 \Pi^{ij}_- (k)= \frac{c_E e^2}{2 \pi^2}  i \epsilon^{i j k} k_k \left( 1- \frac{\omega^2}{|\mathbf{k}|^2}  \right) \left[1-  \omega L(k) \right] \ ,
\ee
where
\be L(k) = \frac{1}{2 |\mathbf{k}|} \log \frac{\omega + |\mathbf{k}|}{\omega - |\mathbf{k}|} \ . \ee

After inspection of the Maxwell equation (\ref{anomalousMaxwell}) for the transverse mode, Laine realized that there are collective unstable modes~\cite{Laine:2005bt}.
Laine also pointed out in Ref.~\cite{Laine:2005bt} that the anomalous HTL effective action could probably be described by some sort of Vlasov equation, in the same way
that the non-anomalous part is~ \cite{Silin,Blaizot:1993zk,Blaizot:1993be,Kelly:1994dh,Kelly:1994ig}.
This is actually so, as we will discuss in the following part of this manuscript.

\section{Kinetic theory with Berry curvature}
 \label{sec:kintheory}

We present in this section a brief review of the (non-covariant)  kinetic theory with Berry curvature valid for  chiral fermions, see 
Refs.~\cite{Son:2012wh,Son:2012zy,Stephanov:2012ki} for a more extended description.

The wavefunction of a chiral fermion picks up a Berry phase when performing a closed path around the point in momentum space ${\bf p}=0$, the point where
the level crossing giving rise to the chiral anomaly takes place. Effects of the Berry phase have to be encoded by changing the particle's Hamiltonian, and modifying its phase-space volume element as well.
In this way, one introduces a fictitious magnetic monopole field in momentum space $\Om$ centered at ${\bf p}=0$.

Once the Berry curvature corrections are considered, the transport equation obeyed by the distribution function $f_p$ of
a fermion carrying charge $e$ is~\cite{Duval:2005vn,Son:2012zy}
\be
\label{transportBerry}
 \frac{\pa f_p}{\pa t} + \frac{1}{1+ e \Bvec \cdot  \Om} \left\{ \left[ e \mathbf{{\tilde E}} + e \mathbf{v} \times \Bvec + e^2 ({\tilde \Evec} \cdot \Bvec ) \Om \right] \cdot \frac{\pa f_p}{\pa \mathbf{p}}
+ \left[ \mathbf{v} + e  {\tilde \Evec} \times \Om + e (\Om \cdot \mathbf{v}) \Bvec \right] \cdot  \frac{\pa f_p}{\pa \mathbf{x}} \right\}= 0  \ , 
\ee
where we have neglected the collision term in the right-hand side,   ${\bf v} = \frac{\partial \epsilon_p}{\partial {\bf p}}$, and $ e {\tilde \Evec} = e \Evec - \partial \epsilon_p/\partial {\bf x}$, and 
the Berry curvature is
\be
\Om = \pm \frac{{\bf p}}{2 |{\bf p}|^3} \ ,
\ee
where the sign $\pm$ corresponds to right- and left-handed fermions, respectively.
The quasiparticle energy is also defined as
\be
\label{mod-dl}
\epsilon_p = \epsilon_p^0 (1 - e  \Bvec \cdot  \Om ) \ ,
\ee
where for massless fermions $ \epsilon_p^0 = |{\bf p}| = p$. We note here that other formulations of the chiral transport equation use a linear dispersion law $\epsilon_p = \epsilon_p^0$.
This also implies that the terms that depend on $\partial \epsilon_p/\partial {\bf x}$  in Eq.~(\ref{transportBerry})  are absent in the transport equation, and thus they lead to a different dynamical evolution of
the fermionic distribution function. The form of this dispersion law has been justified in Ref. \cite{Son:2012zy} at $T=0$. Independently of the value of  $T$ this form of the dispersion law can also be justified after performing a
Foldy-Wouthuysen diagonalization of the Dirac Hamiltonian in the presence of electromagnetic fields in powers of $\hbar$, see the results of Refs.~\cite{Bliokh:2005nk,Gosselin:2006ht} particularized to the massless case.

The particle density is expressed as
\be
n =   \int \frac{d^3 p}{(2 \pi)^3 } (1 + e \Bvec \cdot  \Om ) f_p \ ,
\ee
while the current is given by 
\be 
\label{part-current}
{\bf j} = - \int \frac{d^3 p}{(2\pi)^3 }\left[ \epsilon_p \frac{\pa f_p  }{\pa \mathbf{p}}
+e \Om \cdot  \frac{\pa f_p}{\pa {\bf p}} \epsilon_p \Bvec + \epsilon_p \Om \times \frac{\pa f_p}{\pa \mathbf{x}} \right]
+ \Evec \times {\bf \sigma }
\ ,
\ee
where 
\be
{ \bf \sigma} =  \int \frac{d^3 p}{(2\pi)^3 } \Om f_p \ .
\ee

In the same way one can define the energy-momentum tensor $t^{\mu \nu}$ associated  to the particles. For example, the energy and momentum densities are 
\ba
\label{part-en}
t^{00} & = &  \int \frac{d^3 p}{(2 \pi)^3 } (1 + e \Bvec \cdot  \Om ) \ \epsilon_p f_p \ , \\
\label{part-mom}
t^{0i} & = &  \int \frac{d^3 p}{(2 \pi)^3 } (1 + e \Bvec \cdot  \Om ) \ p^i f_p \ ,
\ea
respectively.

Let us finally stress that in the absence of the Berry curvature terms $\Om = 0$ the transport equation (\ref {transportBerry}) 
reduces to the standard transport equation of a charged particle interacting with electromagnetic fields. 

Using the transport equation (\ref{transportBerry}) it is possible to show that  the particle current obeys 
\be \label{eq:anomaly}
\pa_t n  + \nabla \cdot {\bf j} = - e^2  \int \frac{d^3 p}{(2\pi)^3 }\left( \Om \cdot \frac{\pa f_p}{\pa {\bf p} } \right) \Evec \cdot \Bvec \ .
\ee
The integral above should be performed with care due to the breakdown of the semiclassical description at the singular point $p=0$, where
the level crossing occurs~\cite{Stephanov:2012ki,Wong:2011nt}. This point acts as a source and drain of chiral fermions due to
the chiral anomaly and it is the responsible of a nonzero divergence of the current~\cite{Nielsen:1983rb}. To perform the integral (\ref{eq:anomaly})
one should define a boundary in momentum space containing the singular point. For simplicity, we can take a 2-sphere of radius $R$ centered at ${\bf p}=0$.
In the classical region i.e. $|{\bf p}|>R$ there is no creation or annihilation of fermions. Therefore, the nonzero divergence in
Eq.~(\ref{eq:anomaly}) is due to the flux of fermions that crosses the surface $S^2(R)$~\cite{Stephanov:2012ki,Wong:2011nt}:
\be \pa_t n  + \nabla \cdot {\bf j} = e^2 \int_{S^2(R)} \frac{d{\bf S}}{(2\pi)^3} \cdot \Om \ f_p \  \Evec \cdot \Bvec \ , \ee
where $d{\bf S}$ is the surface element of the 2-sphere. If the distribution function depends only on $p= |{\bf p}|$ the integral is immediate. Then one takes the limit
$R \rightarrow 0$ to reach the singular point, and obtains \cite{Stephanov:2012ki}
\be 
\label{single-anomaly}
\pa_t n  + \nabla \cdot {\bf j} = \pm \frac{e^2}{4\pi^2} \ f_{p=0}  \ \Evec \cdot \Bvec \ . 
\ee
If one considers a theory with both right and left-handed fermions, this equation  reproduces the equation of triangle chiral anomalies appearing in quantum field theory
with a well-defined Fermi surface, as the Fermi-Dirac distribution function is 1 at $p=0$ and $T=0$ \cite{Son:2012wh,Son:2012zy,Stephanov:2012ki}.

At any small but finite value of the temperature $T$, Eq.~(\ref{single-anomaly}) suggests that the chiral anomaly might  get thermal corrections. Of course, it is well known that in relativistic quantum field theories
the chiral anomaly does not receive thermal corrections~\cite{Itoyama:1982up,Liu:1988ke,GomezNicola:1994vq}.
We note here that if we also describe the antifermions with the same sort of transport equation with Berry curvature  one 
obtains the proper chiral anomaly equation of a relativistic quantum field theory at finite $T$. This was mentioned in Ref.~\cite{Stephanov:2012ki}, but let us see how this effectively happens
after taking into account all the quantum numbers associated to the particles/antiparticles.
Noting that at finite temperature each fermion ($f_p^R,f_p^L$) has a corresponding antifermion ($\bar{f}_p^L,\bar{f}_p^R$), 
the divergence of the axial current $j^\mu_A=j^\mu_R-j^\mu_L$ reads 
\be \pa_\mu j_A^\mu =  \frac{e^2}{4\pi^2} \left( f^R_{p=0} + \bar{f}^{L}_{p=0} + f^L_{p=0} + \bar{f}^{R}_{p=0} \right) \ \Evec \cdot \Bvec \ . 
 \ee
 Now, for a system in thermal equilibrium, we have  $f^{R,L}_p=n_F(p-\mu_{R,L})$ and $\bar{f}_p^{R,L} = n_F (p+\mu_{L,R})$, where $n_F (x) = 1/(e^{x/T} +1) $ is the thermal
 Fermi-Dirac distribution function, then we get
 
\be \pa_\mu j_A^\mu =  \frac{e^2}{2\pi^2} \Evec \cdot \Bvec \ , 
 \ee
which shows that the chiral anomaly does not receive thermal corrections.

From this kinetic theory, it is also possible to construct the electromagnetic current, from every species of   particle's current 
\be
J^\mu = (\rho, {\bf J}) = \sum_{s= {\rm species}} e_s \left[  ( n_s, {\bf j}_s) \,  -   (\bar{ n}_s, \bar {{\bf j}}_s) \right]  \ , 
\ee
where $e_s$ is the charge associated to a given species of particles, and
we include both fermions and antifermions with different chiralities of different species.
The electromagnetic current is then used as a source in Maxwell equations
\be
\label{Maxwell}
\pa_\mu F^{\mu \nu} = J^\nu \ .
\ee

Consistency with gauge invariance requires that the electromagnetic current is conserved. Starting with the full kinetic equation
one then has conservation of the current if 
\be
\label{Anomalycancellation}
\sum_{s={\rm species}}   e_s^3  \left \{ \int \frac{d^3 p}{(2\pi)^3 }\left( \Om^s \cdot \frac{\pa f^s_p}{\pa {\bf p} } \right) 
-  \int \frac{d^3 p}{(2\pi)^3 }\left( \Om^s \cdot \frac{\pa \bar{f}^s_p}{\pa {\bf p} } \right) \right \}
= 0 \ .
\ee
We note that in a theory in thermal equilibrium with both left- and right-handed fermions and antifermions, the gauge anomaly cancels, regardless
of the value of their associated chemical potentials. This is so, as for a system in thermal equilibrium Eq.~(\ref{Anomalycancellation}) reads 
\be
\label{Anomalycancellation2}
 - \frac{e^3}{4 \pi^2}  \left( f^R_{p=0} + \bar{f}^{L}_{p=0} - f^L_{p=0} - \bar{f}^{R}_{p=0} \right) = 0 \ .
\ee
Thus, we see that if there is not a gauge anomaly in the theory at $T=\mu =0$, there is not a gauge anomaly
at any finite value of the temperature and density.

While it would be interesting to derive the kinetic equation obeyed by the chiral antifermions from the underlying quantum field theory, we will not do it here.
We will simply check that once antiparticles are taken into account on the same footing as particles, the kinetic equation with Berry curvature not only reproduces
correctly the chiral anomaly equation. It also gives proper account of the anomalous gauge polarization tensors described in Sec.~\ref {aHTLeffectiveaction}.
We do so in the following section.

\section{Linear Response Analysis\label{sec:lrt}}

Here we use linear response theory to prove that the kinetic theory with Berry curvature presented in the
previous Section reproduces the anomalous HTL effective action of Sec.~\ref{aHTLeffectiveaction}. The fact that it reproduces the results
of quantum field theories with a chemical potential assigned to chiral fermions at zero temperature was already
checked in Ref.~\cite{Son:2012zy}. Here we see that it also matches results obtained at high temperature.

Let us first recall that it has been known for a while that the HTL (or HDL) polarization tensor in an Abelian \cite{Silin} and also in a  non-Abelian plasma
\cite{Blaizot:1993zk,Blaizot:1993be,Kelly:1994dh,Kelly:1994ig}  can be reproduced
from transport theory. Here we show that  new terms in the transport theory associated to the Berry curvature can also reproduce the anomalous HTL.
In order to check that statement, let us set the same power counting analysis of Ref.~\cite{Son:2012zy}. We assume weak fields, and that
the vector gauge field is of order
 $A_\mu = \cal{O} (\epsilon)$, while the coordinate derivatives are 
$\partial/\partial x^\mu= \cal{ O} (\delta )$. We will compute the
electromagnetic particle current up to the order  ${\cal O} (\epsilon \delta )$ for isotropic configurations.

Assuming a known initial isotropic distribution, we study the evolution of the system under a small perturbation. Then the distribution functions is 
expanded  as
\be f_p (\epsilon_p,{\bf p}) = f_p^{(0)}(\epsilon_p,{\bf p}) + e\left( f_p^{(\epsilon)}(\epsilon_p,{\bf p}) + f_p^{(\epsilon \delta)} (\epsilon_p,{\bf p})\right) + \cdots \ee
However, because $\epsilon_p$ has also a ${\bf B}$ dependence,  to organize properly the power counting, 
one should take into account that
\be
 f_p^{(0)}(\epsilon_p,{\bf p}) =   f_p^{(0)}(\epsilon^0_p,{\bf p}) - e
  \Bvec \cdot  \Om \frac{ d f_p^{(0)}(\epsilon^0_p,{\bf p})}{d \epsilon^0_p} + \cdots  \ee
and express all the distributions functions as a function of $\epsilon^0_p$. Thus, we will organize the expansion around $\epsilon^0_p$, although we will not
write explicitly such a dependence in what follows.
Further, we consider massless fermions, and thus $\epsilon_p^0 = |{\bf p}|$.

From Eq.~(\ref{transportBerry}) it is easy to obtain the equations obeyed by every order in the approximation.
At first order
\be (v \cdot \pa) \ f_p^{(0)} = 0 \ ,
\ee
where $v \cdot \pa = \pa_t + {\bf v}\cdot \pa_{\bf x}$.
At order $\cal{O} (\epsilon)$  the distribution function obeys
\be  
\label{line-e}
(v \cdot \pa) \ f_p^{(\epsilon)} =  -  \mathbf{E} \cdot \mathbf{v} \ \frac{df_p^{(0)}}{dp} \ ,
\ee
while at order  $\cal{O} (\epsilon \delta)$ one has
\be  
\label{line-ed}
(v \cdot \pa) \ f_p^{(\epsilon \delta)} = \pm \frac{1}{2p} \ \mathbf{v}\cdot \pa_t \Bvec \ \frac{df_p^{(0)}}{dp} \ ,
\ee
with $\pm$ for right-handed and left-handed fermions, respectively.
 We note here that  the formulations of the chiral transport equation that use a linear dispersion law
instead of Eq.~(\ref{mod-dl}) lead to a different solution for $f$ at this order, as then the right-hand side of Eq.~(\ref{line-ed}) would vanish.
In particular, this would lead to different values of the associated current at this order, and in turn,  to a different value
of   parity odd part of   the photon polarization tensor
 
 We note here that  the formulations of the chiral transport equation that use a linear dispersion law
instead of Eq.~(\ref{mod-dl}) lead to a different solution for $f$ at this order, as then the r.h.s. of Eq.~(\ref{line-ed}) would vanish.
In particular, this would lead to different values of the associated current at this order, and in turn,  to a different value
of   parity odd part of   the photon polarization tensor

It is then easy to obtain the electromagnetic current associated to  that particle.
At zero order we do not compute the contribution, as we assume charge neutrality.
At order $\mathcal{O} (\epsilon)$ 
\be
J^{\mu (\epsilon)} =  e^2  \int \frac{d^3 p}{(2\pi)^3 } v^\mu  f_p^{(\epsilon)} \ ,
\ee
which can be expressed, after solving Eq.~(\ref{line-e}) in Fourier space, as 
\be J^{\mu (\epsilon)}  (k)= \left( e^2 \int \frac{dp p^2}{2\pi^2 }   \frac{df_p^{(0)}}{dp} \right) \int_v v^\mu \left[ \delta^{\nu 0} A_\nu - \omega \frac{v^\nu A_\nu}{v\cdot k } \right] \ ,
\ee

At order $\mathcal{O} (\epsilon \delta)$ there is no contribution to $J^0$.
To compute the contribution of  $\mathcal{O} (\epsilon \delta)$, we have to consider the expression of the current Eq.~(\ref{part-current}), and thus
\be J^{i(\epsilon \delta)} = e^2  \int \frac{d^3 p}{(2\pi)^3 } \left( v^i f_p^{(\epsilon \delta)} 
\mp \frac{df_p^{(0)}}{dp} \frac{B^i}{2p} \pm \epsilon^{ijk} \frac{v^j}{2p}\frac{ \pa f_p^{(\epsilon)}}{\pa x^k} \right) \ .
 \ee
Solving the Eqs.~(\ref{line-e}) and (\ref{line-ed})
 for $n^{(\epsilon)}$ and $n^{(\epsilon \delta)}$, respectively, the current at order $\mathcal{O} (\epsilon \delta)$ can then be expressed as

\be J^{i(\epsilon \delta)}(k) = \pm i e^2 \left(- \int \frac{dpp}{4\pi^2}  \frac{df_p^{(0)}}{dp} \right) \left\{\epsilon^{ijk} k^k + \omega 
  \int_v\frac{ (   v^i  \epsilon^{jkl} - \epsilon^{ikl}  v^j) v^k k^l }{v \cdot k } \right\} A_j
  \ .
  \ee
  
Note that at this order fermions with different chiralities contribute with different sign to the current.
Thus, if the distribution function of left- and right-handed fermions is the same, the contribution  to the current at this order vanishes.

To make contact with the theory discussed in  Sec.~\ref{aHTLeffectiveaction}, let us assume we have the same particle content.
This means that in the plasma we consider right-handed and left-handed fermions with charge $e$ and chemical potential $\mu_R$ and $\mu_L$, respectively,
described by the Fermi-Dirac distribution function $n_F(x)$. We should consider also left-handed/right-handed antifermions, carrying charge $-e$ and
with chemical potential $-\mu_R/\mu_L$.
It is then easy to check that the total electromagnetic current is, in Fourier space,
\be
\label{k-elcurrent}
J^\mu (k)=  \Pi^{\mu \nu}_+ (k) A_\nu (k) + \Pi^{\mu \nu}_- (k) A_\nu (k) \ ,
\ee
where $\Pi^{\mu \nu}_+$ and $\Pi^{\mu \nu}_-$ are the polarization tensors expressed in Eqs.~(\ref{pol-HTL}) and (\ref{pol-aHTL}), respectively, and where 
the parameters present in the current are
\ba
m^2_D &= & - e^2  \int \frac{dp p^2}{2 \pi^2}  \left[ \frac{dn_F(p-\mu_R)}{dp} + \frac{dn_F (p+\mu_R)}{dp} 
+ \frac{dn_F(p-\mu_L)}{dp} + \frac{dn_F (p+\mu_L)}{dp}\right ]\ , \label{eq:mD}\\
c_E & =&  \frac 12 \int dp p  \left[ \frac{dn_F (p- \mu_R)}{dp} -\frac{dn_F (p+ \mu_R)}{dp} - \frac{dn_F(p-\mu_L)}{dp} + \frac{dn_F (p+\mu_L)}{dp} \right ]  \label{eq:cE} \ ,
\ea
which results in
\be
m^2_D =  e^2 \left(\frac{T^2}{3} + \frac{\mu_R^2+\mu_L^2}{2 \pi^2} \right) \ ,   \qquad c_E  =  - \frac{\mu_5}{2}\ . \label{eq:axichem}
\ee

With the knowledge of the current, one can obtain an associated effective action by solving
$J^\mu (x) = - \frac{\delta {\cal S}}{\delta A_\mu (x)}$. 
In this way, one can prove that the kinetic theory with Berry curvature allows to describe properly the anomalous terms obtained in a
quantum field theory when a finite chemical potential is assigned to chiral fermions, not only at $T=0$~\cite{Son:2012zy}, but also when the temperature $T$ is high.

Let us finally remark that after analyzing the dispersion law obtained with this classical kinetic theory of the collective gauge field modes 
it was found out that there are modes which are unstable~\cite{Akamatsu:2013pjd}, in full agreement with the observation of Ref.~\cite{Laine:2005bt}.
The instability is a proof that a configuration where there is an imbalance of left- and right-handed fermions cannot be an equilibrium configuration.
The system then evolves dynamically so as to erase such an imbalance.

\section{Energy and momentum densities of the chiral plasma collective excitations\label{sec:hamiltonian}}

As we have seen in Sec.~\ref{aHTLeffectiveaction},  the effective action describing the soft gauge fields  in a chiral plasma is nonlocal.
However, one can recast it in local form by introducing new variables, the distribution functions in coordinate and momenta of the
hard fermionic degrees of freedom. One could wonder also whether the collective gauge field modes described by such a non-local
effective action has a simple expression for their energy and momentum densities. Again, with the use of kinetic theory, one can
write an energy density for those collective modes in a local form, as we show in this section.  Here we simply 
 generalize the procedure used in Ref.~\cite{Blaizot:1994am} to construct the energy-momentum tensor
for the collective modes of a  quark-gluon plasma.

In the kinetic theory framework, one defines the total energy-momentum tensor of the plasma 
\be
T^{\mu \nu} (x) = \Theta^{\mu \nu} (x) + t^{\mu \nu} (x) \ ,
\ee
as the sum of the pure gauge field part
\be
\Theta^{\mu \nu } (x) = \frac 14 \, \eta^{\mu \nu}  F_{\rho \sigma}(x) F^{\rho \sigma} (x) - F^{\mu}_{\ \alpha}(x) F^{\alpha \nu} (x) \ , 
\ee
and the part stored in the particles, $t^{\mu \nu}$.  Expressions of the energy density $t^{00}$ and momentum density $t^{0i}$
were given in Eqs.~(\ref{part-en}, \ref{part-mom}).

While the total energy-momentum tensor is conserved, $\partial_\mu T^{\mu \nu} =0$, one has
\be
\label{conser-parttensor}
 \pa_\nu t^{\mu \nu} (x) = F^{ \mu \nu} (x) J_\nu (x) \ .
\ee
In linear response theory, we have seen that the electromagnetic current becomes a non-local function of the gauge fields.
The same should occur for the energy-momentum tensor $t^{\mu \nu}$. However, there is a way to express  it in local form.

Let us define two functions $W_+(x;v)$ and $W_-(x;v)$ obeying the equations
\ba
\label{eqsW+}
 (v \cdot \pa) \  W_+(x;v) & = & {\bf v} \cdot {\bf E}(x) \ , \\
 \label{eqsW-}
 (v \cdot \pa) \ W_-(x;v) & = & -{\bf v} \cdot \pa_t {\bf B}(x) = {\bf v} \cdot \left( \nabla \times {\bf E}(x) \right) \ ,
  \ea
respectively. Note that in the last equation we have used the Faraday's law $ \pa_t {\bf B} +  \left( \nabla \times {\bf E} \right) =0$.

Combining Eqs.~(\ref{eqsW+},\ref{eqsW-}) with Eqs.~(\ref{line-e},\ref{line-ed}) it is possible to obtain the relation between $W_{\pm}$ and the functions $f_p^{(\epsilon)}$ and $f_p^{(\epsilon \delta)}$,
\ba m_D^2 \ W_+ (x;v) &=& e^2 \int \frac{dp p^2}{2\pi^2} \sum_{s,\lambda} \ f_p^{(\epsilon)} (x,p,v) \ , \\
\frac{c_E}{2\pi^2} \ W_- (x;v) &=& - \int \frac{dp p^2}{2\pi^2} \sum_{s,\lambda} \ f_p^{(\epsilon \delta)} (x,p,v) \ , \ea
where the sum runs over species and helicities in the same fashion as Eqs.~(\ref{eq:mD},\ref{eq:cE}).

The electromagnetic current Eq.~(\ref{k-elcurrent}) can then be expressed in terms of these functionals as follows
\be
J^\mu (x) =  m^2_D \int_v v^\mu \ W_+(x;v) - \frac{c_E e^2}{2 \pi^2} \delta^{\mu i} \int_v \left \{ v^i \  W_-(x;v) +  B^i(x) - [ {\bf v}\times \nabla W_+(x;v)]^i \right \} \ .
\ee
In the static limit $\omega \rightarrow 0$ one immediately gets the CME current
\be 
{\bf J} (x)=  \  \frac{e^2 \mu_5}{4 \pi^2} \ {\bf B} (x) \ ,
\ee
where Eq.~(\ref{eq:axichem}) has been used.

Let us consider the zeroth component of Eq.~(\ref{conser-parttensor}). The left-hand side of the equation can then be expressed as
follows
\ba
\label{Excurrent}
{\bf E}(x) \cdot {\bf J} (x) &=& m^2_D \int_v  {\bf E}(x) \cdot {\bf v} \ W_+(x;v) \\
&+& \frac{c_E e^2}{2 \pi^2}\int_v 
\left[  - {\bf E}(x) \cdot {\bf v}\, W_-(x;v) - {\bf E}(x) \cdot {\bf B}(x) + {\bf E}(x) \cdot ( {\bf v}\times \nabla W_+(x;v)) \right] \ . \nonumber
\ea

We integrate by parts the last term of Eq.~(\ref{Excurrent}), and use Eqs.~(\ref{eqsW+}) and~(\ref{eqsW-})
 to find
\ba
{\bf E}(x) \cdot {\bf J} (x) &= & m^2_D \int_v  \ W_+(x;v) \  v \cdot \pa \ W_+(x;v)  \\
& - & \frac{c_E e^2}{2 \pi^2} \int_v  \left[ \ W_-(x;v) \ v \cdot \pa \  W_+(x;v)  +  W_+(x;v) \ v \cdot \pa \ W_-(x;v) \ \right] 
 - \frac{e^2 c_E}{2 \pi^2}  {\bf E}(x) \cdot {\bf B} (x) \ . \nonumber
\ea
The Lorentz invariant ${\bf E} \cdot {\bf B}$ can be expressed as a total derivative
\be
 {\bf E}(x) \cdot {\bf B} (x) = \frac 14 F_{\mu \nu} (x) {\tilde F}^{\mu \nu} (x) = \frac14\partial_\mu J^\mu_{\rm CS} (x) \ ,
\ee
where 
\be
J^\mu_{\rm CS}(x) = \epsilon^{\mu \nu \rho \sigma} A_\nu(x) F_{\rho \sigma} (x)
\ee
is the Chern-Simons current. We can then write
\be
\pa_\mu t^{0 \mu}(x) =\pa_\mu \left(   \frac{m^2_D}{2}  \int_v  v^\mu \  W_+(x;v) \,W_+(x;v) - \frac{c_E e^2}{2 \pi^2}  \int_v v^\mu \ W_+(x;v) \,W_-(x;v) -\frac{c_E e^2}{8 \pi^2} J^\mu_{\rm CS}(x) \right) \ .
\ee
This means that, up to a constant, we can identify the  values of  $t^{00}$ and $t^{0i}$. Because $W_+$ and $W_-$ represent the contribution to the current at order $\cal{O}(\epsilon)$ and 
$\cal{O}(\epsilon \delta)$, respectively, we fix the value of the constant by considering the contribution to $t^{00}$ and $t^{0i}$ at zeroth order. Then  

\be
t^{00} (x) =  \epsilon^{(0)} + \frac{m^2_D}{2}  \int_v   W_+(x;v) \,W_+(x;v) - \frac{c_E e^2}{2 \pi^2}  \int_v  W_+(x;v) \,W_-(x;v) - \frac{c_E e^2}{8 \pi^2}  J^0_{\rm CS} (x)\ ,
\ee
and
\be
t^{0i} (x)=  \frac{m^2_D}{2}  \int_v  v^i \ W_+(x;v) \,W_+(x;v) - \frac{c_E e^2}{2 \pi^2} \int_v v^i \ W_+(x;v) \,W_-(x;v) - \frac{c_E e^2}{8 \pi^2}  J^i_{\rm CS} (x) \ ,
\ee
where the energy density $\epsilon^{(0)}$ is associated to the $n_p^{(0)}$ distribution function. Assuming the presence of both right- and left-handed chiral fermions
\ba \label{eq:freeen}
\epsilon^{(0)} & = & \int \frac{dp p}{2\pi^2} \int_v \epsilon_p^0 \ \left[ \ n_F (p - \mu_R) + n_F(p + \mu_R) +
n_F (p - \mu_L) + n_F(p + \mu_L) \ \right] \\
&  = & \frac{7 \pi^2 T^4}{60} + \frac{\mu_R^2 +\mu_L^2 }{4}T^2 + \frac{\mu_R^4 + \mu_L^4}{8 \pi^2} \ . \nonumber
\ea

The energy of the total system can then be expressed as
\be
{\cal E} = \int d^3 x \ \left[ \ \Theta^{00}(x) + t^{00}(x) \ \right] \ .
\ee 
In the $A_0 =0$ gauge it reads
\ba
\label{lastH}
{\cal E} & = &  \int d^3 x \left[ \frac 12 {\bf E}^2(x) + \frac 12 {\bf B}^2(x) + \epsilon^{(0)}+ \frac{m^2_D}{2}   \int_v   W_+(x;v) \,W_+(x;v) \right. \\
 & - &\left. \frac{c_E e^2}{2 \pi^2}  \int_v  W_+(x;v) \,W_-(x;v) +\frac{c_E e^2}{4 \pi^2}  {\bf A}(x)\cdot {\bf B}(x) \right] \ . \nonumber
\ea

The four first terms are the same terms as in the energy density for the collective modes in the strict (non-anomalous) Abelian HTL theory \cite{Blaizot:1994am}. 
The last two pieces are the new terms arising due to the Berry
curvature corrections. We note that the last piece in Eq.~(\ref{lastH}) corresponds to a term proportional to the magnetic helicity, defined as \cite{biskamp}
\be
H_M = \int_V d^3 x \, {\bf A}(x)\cdot {\bf B}(x) \ .
\ee
This quantity is gauge invariant provided the normal component of ${\bf B}$  to the surface of the volume under integration vanishes or the magnetic field itself cancels at the boundary $\pa V$.

Notice that the first four terms  of Eq.~(\ref{lastH}) are strictly positive, the last two terms are not manifestly positive. Because  they represent 
corrections to the dominant term $m_D^2$ in the high $T$ (or high $\mu$)  limit, suppressed by an order $\partial/\partial x^\mu \sim {\cal O} (\delta)$, we expect them to be subdominant, as long as one
only considers long wavelength fields and weak fields. Thus, the total energy density of the system should be positive.

It should be possible to analyze the Hamiltonian dynamics and Poisson bracket structure of the present theory. Such a study was already performed by Nair~\cite{Nair:1994xs} (see Ref.~\cite{Blaizot:2001nr} for a review) for the case where the anomalous
contributions are absent. We leave a similar study for the Hamiltonian dynamics in presence of anomalous terms for a future publication.

\subsection{Static configurations and energy content}
\label{sec:CSwave} 

The spatial components describe the so-called Chern-Simons wave~\cite{Rubakov:1985nk,Rubakov:1986am}. In an orthonormal Cartesian basis with the first component
in the direction of ${\bf k}$ we have
\ba A^1(x) & = & 0 \\
    A^2 (x) & = & |A| \sin \ (k_{CS} \ \hat{\bf{ k}} \cdot {\bf r} +b ) \\
  A^3(x) &=&  |A| \cos \ (k_{CS} \hat{\bf{ k}} \cdot {\bf r} + b) \ ,
\ea
with $|A|$ and $b$ constants of integration, $\hat{{\bf k}}={\bf k}/k$ and
\be k_{CS} = \frac{|c_E| e^2}{2\pi^2} \ .
 \ee
 
In the temporal gauge $A_0 =0 $ the energy of the static system, consisting of the Chern-Simons wave reads 
\be
 \label{eq:hamilstatic}
{\cal E} = \epsilon^{(0)} V + k_{CS}^2 |A|^2 V \ , \ee
where $V$ is the volume of the system.

After analyzing the dynamical evolution of the Chern-Simons wave, one can detect the existence of an unstable gauge mode~\cite{Laine:2005bt}.
The energy associated to that gauge mode would grow, and this is only achieved if the energy associated to the fermionic degrees of freedom 
diminishes. We leave the specific study of such a dynamical evolution for a future project.

\section{Conclusions and Outlook~\label{sec:conclusions}}

We have revised the recently proposed kinetic theory with Berry curvature corrections and checked that also at finite temperature it describes the
same quantum chiral anomaly equation of quantum field theories, when both fermions and antifermions are considered. Thus, the domain
of application of the transport approach goes beyond those systems with well-defined Fermi surfaces, as claimed in most part of the literature. We have also checked that the transport approach of Ref.~\cite{Son:2012zy}
describes the anomalous HTL/HDL  Feynman diagrams that appear in Abelian gauge theories with a chirality fermion imbalance.
 It is worth emphasizing here that different formulations of the relativistic chiral transport equation differ in the form of  fermionic dispersion law and in explicit terms
of the kinetic equation, and lead to different forms of the photon polarization tensor.
Our results can be viewed  as a consistency check of the transport equation of Ref.~\cite{Son:2012zy}. A formal derivation of
that transport approach for both particles and antiparticles for any value of $T$ or $\mu$ would be desirable.

Transport theory is known to be a sort of effective theory to describe the HTL/HDL action, and it seems that the  same occurs for the effective
action associated to the anomalous Feynman diagrams which occur in chiral plasmas. The use of transport theory versus a quantum field theory approach shows to be more versatile to study pure non-equilibrium phenomena. It is also
 quite convenient to study  the dynamical evolution of the plasma, both close or far away from equilibrium.
 In the presence of a chirality imbalance, represented by the chemical potential $\mu_5 = \mu_R -\mu_L$, 
the two formalisms show the existence of an unstable collective gauge field \cite{Laine:2005bt,Akamatsu:2013pjd}.  This means that there are gauge field modes that would grow
exponentially, until the instability is saturated, and the chirality imbalance is washed out. In this manuscript we have constructed an energy density, valid in the limit of small couplings or also weak fields, given in local and quadratic form. Studying the time evolution of the gauge fields and the $W_+$ and $W_-$ distribution functions, together with the anomaly equation should be enough to see how the chiral plasma instability is saturated.  We stress here that the Hamiltonian for the HTL collective modes has already been successfully used for numerical simulations (see, e.g. Ref.~\cite{Bodeker:1999gx}), and a variant of it  was also used to study how the Weibel instabilities were saturated for anisotropic non-Abelian plasmas \cite{Arnold:2005vb,Rebhan:2004ur}. Similarly, the formulation presented here can be used for the numerical studies  of how the CME evolves dynamically. We leave such a study for a future project.

\acknowledgments{
 This research was supported by the Spanish MINECO,  under contract FPA2010-16963. The work of J.M.T.-R. is funded by Grant No. FP7-PEOPLE-2011-CIG under
Contract No. PCIG09-GA-2011-291679.

\end{document}